\documentclass{ws-procs9x6}

\begin{document}

\title{Exotic superfluidity in cold atoms}

\author{Elena Gubankova \footnote
{This work is done in collaboration with Massimo Mannarelli and Rishi Sharma}}

\address{Tufts University, Boston, MA\\
E-mail: elena1@mit.edu}

\begin{abstract}
We derived the low energy effective action for the collective modes in asymmetric fermionic
systems with attractive interaction. We obtained the phase diagram in terms of the chemical potentials.
It features a stable gapless superfluidity with one Fermi surface on the BEC side of the resonance. Also
we predict a sharp increase in outer core of a vortex, i.e. vortex size, upon entering into the gapless phase.
This may serve as a signature of a gapless phase.  

\end{abstract}

\keywords{BCS-BEC crossover; asymmetric ultracold Fermi systems; gapless superfluidity.}

\bodymatter

\section{Introduction}
Experiments in cold atomic gasses renewed the interest in fermionic pairing \cite{ketterle,Partridge}.
In particular, much effort was made to understand superfluidity in imbalanced
fermionic gasses. Imbalanced systems have unequal number of particles of two fermionic
species that pair. The system consists of fermions of two different species, which correspond to two
hyperfine states of a fermionic atom such as $^6{\rm Li}$. These fermions have opposite spin and interaction
between them can be tuned by using a Feshbach resonance \cite{Feshbach}. 
   
For zero imbalance, the system properties are qualitatively well understood using mean field \cite{BEC}.
In weak coupling the system lives in a weakly coupled BCS state and crosses over to a strongly coupled
BEC state through the resonance region. While the extreme BCS and BEC regimes are in good control
in mean field theory, quantitative understanding of the phases close to resonance comes mainly from
Monte-Carlo calculations \cite{Carlson:2005kg}. At resonance, fluctuations change mean field only quantitatively.
Situation is completely different at nonzero polarization, where fluctuations change
mean field results around the resonance qualitatively; and many features of the phase diagram are not 
caught by the mean field. For imbalanced systems, 
a qualitatively complete picture of the phase diagram has not been established yet.
At small asymmetry, and weak coupling,
there is standard BCS superfluidity, which breaks down at Clogston limit. At large asymmetry,
proposed possibilities are phase-separation \cite{Bedaque:2003hi}, gapless (or breached) superfluidity \cite{Sarma,Liu:2002gi,Gubankova:2003uj,Forbes:2004cr},
deformed Fermi sea pairing \cite{Muther:2002mc}, and non-homogeneous or LOFF pairing \cite{LOFF}.
 
The phase diagram of the system at T=0 as a function of scattering length and the chemical potential
difference has been explored. In the mean field approximation \cite{Sheehy:2006qc},
it was found that on the BCS side
of the resonance there are no stable gapless superfluid phases with net polarization; and at the resonance
the first order phase transition from superfluid to normal phase takes place as asymmetry is increased,
without any intervening gapless superfluids. In experiment with trapped atoms this leads that the system
phase separates, because the gapped phase can not feature a net polarization, unpolarized 
superfluid is in the central region of the trap and a polarized normal fluid is at the exterior \cite{Sheehy:2006qc,trap}.    
There was an effort done to go beyond the mean field approximation by using Monte-Carlo simulations \cite{Carlson:2007},
an expansion in $\epsilon=d-4$ space dimensions \cite{Son:2005qx}, 
$1/N$ loop expansion around the BCS-BEC solution \cite{Nikolic:2007,Radzihovsky:2007},
a superfluid local density approximation \cite{Bulgac:2006cv,Bulgac:2007wm}.

The authors of~\cite{Son:2005qx} propose a phase diagram
in the plane of scaled polarization $\eta\sim\delta\mu$ 
and diluteness parameter $\kappa=-1/na^3$ which features a splitting point near the resonance 
at nonzero mismatch
where the homogeneous superfluid, a LOFF like inhomogeneous phase, 
and the gapless superfluid phase coexist. They also find
stable gapless fermionic modes with one and two Fermi surfaces on the BCS side of the resonance.
This picture is supported by Monte-Carlo simulations \cite{Carlson:2007}.

In order to understand how fluctuations in the condensate around the mean field value affect
the phase diagram of cold atomic gasses with asymmetry, we study the effective theory describing 
these fluctuations.

\section{Effective low energy theory}
We consider a non-relativistic system consisting of two species of fermions $\psi_1$ and $\psi_2$
of equal mass $m$  but different chemical potentials $\mu_1$ and $\mu_2$. We assume that the Feshbach
interaction between fermions of different species, given via the scattering length,
can be modeled by a point like four Fermi interaction with coupling $g = 1/(k_F a_s)$,
$n=k_F^3/(2\pi^2)$. BCS is at $g\rightarrow -\infty$, $a_s < 0$; BEC is at $g\rightarrow\infty$, $a_s > 0$.
The effect of the attractive interaction
between fermions is to produce a difermion condensate. Condensate spontaneously breaks $U(1)$ associated
with conservation of total fermion number (fermions are not charged); hence there is one Goldstone mode
and the system is superfluid. By gauging $U(1)$, condensate spontaneously breaks local $U(1)$;
hence the gauge boson becomes massive, i.e. there is Meissner mass and the system is a superconductor.
We established a correspondence between coefficients of effective theory for Goldstone mode and screening
masses of the gauge field. Mean field Lagrangian is quadratic in fermion fields,
\begin{equation}
{\cal{L}} = \Psi^\dagger \left(
\begin{array}{cc}
i{\partial_t}-\xi({\bf p})+\delta\mu\sigma^3 & -\Delta(x)\varepsilon    \\
\Delta^*(x)\varepsilon &   i{\partial_t}+\xi({\bf p})-\delta\mu\sigma^3
\end{array} \right)\Psi -\frac{|\Delta(x)|^2}{\lambda}
\end{equation}
$\Psi$ stands for the four component Nambu-Gorkov spinor. We consider
a homogeneous condensate, independent of x, i.e. $\Delta(x)=\Delta$, as a mean-field solution. 
The quasiparticle dispersion law is $\epsilon_{\pm} =\pm\delta\mu + \sqrt{{\xi({\bf p})}^2+\Delta^2}$.
At some momenta it features zero energies, describing spherical Fermi surfaces in momentum space.

We include fluctuations of the condensate around the mean field solution,
by introducing the field $\eta(x)$, $\Delta(x) = \Delta+\eta(x)$. Partition function contains fermion field $\Psi$ and 
fluctuation of the condensate fields,
\begin{equation}
Z=\int{\cal{D}}\eta^*{\cal{D}}\eta{\cal{D}}\Psi^\dagger{\cal{D}}\Psi
e^{-{\cal{S}}[\Psi^\dagger,\Psi,\eta,\eta^*]}
\end{equation}
 To get an effective action for the fluctuations, we integrate
out the fermion fields; we get the fermion determinator,
\begin{eqnarray}
S[\eta,\eta^*] &=& \int d^4x\Bigl\{\frac{1}{g}|\Delta+\eta(x)|^2\Bigr\}\nonumber\\
&-& \Bigl\{\frac{1}{2}{\rm{Tr}}\log \left(
\begin{array}{cc}
{-\partial_{x^4}}-{\xi({\bf p})}+\delta\mu & -(\Delta+\eta(x))    \\
-(\Delta+\eta^*(x))  &   {-\partial_{x^4}}+{\xi({\bf p})}+\delta\mu
\end{array} \right)\Bigr\}
\end{eqnarray}
We expand the fermion determinator
in fluctuations and their derivatives, ${\cal{S}}[\eta,\eta^*]={\cal{S}}^{(0)}+{\cal{S}}^{(1)}+{\cal{S}}^{(2)}+...$~.
As a result we get an effective theory of collective modes
(Goldstone and Higgs) away from unitarity. To the leading order, $S^{(0)}$ is the free energy of the system
without fluctuations. Due to the gap equation the next term $S^{(1)}$ vanishes 
(stationary point of the action). The lowest nontrivial term in the expansion is $S^{(2)}$,
\begin{equation}
{\cal{S}}^{(2)}=
-\frac{1}{2}\frac{T}{V}\sum_k \left(\lambda(-k)\theta(-k)\right)
\left( \begin{array}{cc}
M_{11}(k) & M_{12}(k)    \\
M_{21}(k) & M_{22}(k)
\end{array} \right)
\left(\begin{array}{cc}\lambda(k) \\ \theta(k) \end{array} \right)
\end{equation}
with $\eta=(\lambda+i\theta)/\sqrt{2}$.
It is UV finite; UV divergent contributions cancel exactly by employing the gap equation.
(Note, the gap equation is UV divergent and it is regulated by the cutoff $\Lambda$, usually 
the Fermi momentum, which is the highest scale in the problem).
We separate magnitude and the phase of the fluctuations, $\eta=(\lambda+i\theta)/\sqrt{2}$. 
The physical meaning of the real and complex
components of the fluctuation field is easy to understand  in small fluctuation and long wavelength
limit,
\begin{equation}
{\cal{S}}^{(2)}\rightarrow
\left( \begin{array}{cc}
-C+Dk_0^2-\frac{E}{3}k^2 & ik_0F    \\
-ik_0F & Ak_0^2-\frac{B}{3}k^2
\end{array} \right)
\end{equation}

Let us first consider the phase of condensate, but not its magnitude, fluctuates, 
$\Delta \rightarrow \Delta e^{i\phi(x)}$. The field $\phi$ 
represents the Goldstone mode associated with the spontaneous breaking of the total fermion number,
$n_1+n_2$. There is no explicit breaking, hence the mass of Goldstone boson is zero. In the low energy
Lagrangian density, 
\begin{equation}
{\cal{L}}_\phi = {\Delta^2} \bigl[A(\partial_t\phi(x))^2 -
\frac{B}{3}(\partial_i\phi(x))^2\bigr]\label{Goldstone}
\end{equation}
$A$ is the kinetic energy, $B$ is the spatial variation of the Goldstone mode.
Negative $A$ or $B$ means instability towards the growth of phase fluctuations. If $A$ and $B$ are positive,
the system is stable and the speed of sound is $\sqrt{B/3A}$. In our analyses we find $A>0$ always, 
it can be $B<0$.

The case when magnitude fluctuates corresponds to the Higgs mode, $\Delta\rightarrow\Delta+\lambda(x)/\sqrt{2}$,
with the Lagrangian of a massive bosonic field
\begin{equation}
{\cal{L}}_\lambda = -\frac{1}{2} C\lambda(x)^2 + \frac{1}{2} D (\partial_t\lambda(x))^2 -
\frac{E}{6}(\partial_i\lambda(x))^2 \label{Higgs}
\end{equation}
with all positive coefficients, $C$, $D$, $E>0$. 
$m_H^2=C/D$ is the mass squared, squared of the gap
in the excitation spectrum. $C \lambda^2$ is change in free energy $S^{(0)}$ due to change in magnitude 
of condensate; $C$ is proportional to the curvature of the potential, $C>0$ is local minimum (stable/meta-stable) and 
$C<0$ is local maximum (unstable). When $D<0$ or $E<0$, mean field solution is unstable with respect to time or
space-dependent fluctuations of the magnitude; $D>0$ always.

$B<0$, $E<0$ mean growth of spatially non-uniform fluctuations, unstable to phase and magnitude
modulation of the condensate. However, these instabilities
point to different ground state of the condensate. $B<0$: condensate develops phase modulation
which carries a current, balanced by a counter-propagating current carried by gapless fermions.
$E<0$: formation of a spatial modulation in the magnitude of the condensate, which does not carry a current.

$F$ mixes Higgs and Goldstone components. $F$ vanishes in weak coupling.

\begin{center}
\begin{figure}[h]
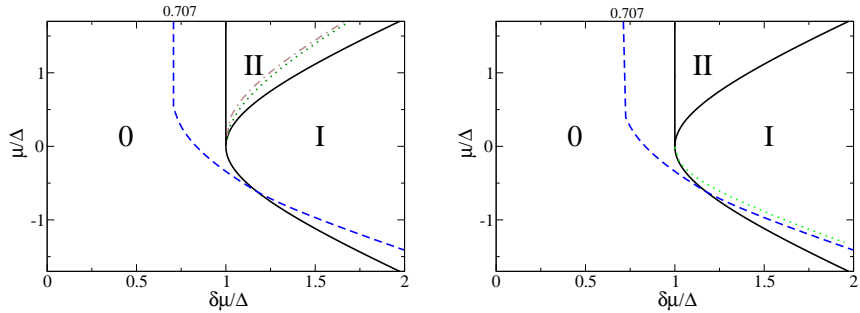

\includegraphics[width=2.1 in,angle=0]{mugrad.eps} \hspace{0.4cm}
\includegraphics[width=2.1 in,angle=0]{mupotlpressure.eps}
\caption{Left panel. The ($\delta\mu,\mu$) plane is divided into three regions
corresponding to different number of gapless surfaces in momentum space, marked 0, I, II (solid lines).
Unstable regions according to criteria 1 and 3: speed of sound for Goldstone and Higgs modes
should be real (local criteria), free energy of the superfluid phase is lower than the free energy
of the normal state (global criteria; $0.707=1/\sqrt{2}$ is Chandrasekhar-Clogston limit of breaking BCS 
superfluidity at weak coupling). Right panel. Unstable regions according to criteria 2 and 3:
superfluid solution should be a minimum
of the free energy (local criteria), pressure of the superfluid state is higher than the pressure
of normal state (global criteria). Requirement 2 excludes states above doted line.  
Criteria 2 is the strongest among local criteria, and it leaves gapless states only with 
one Fermi surface. Local minimum of free energy ($C>0$) is equivalent that the number susceptiblity
matrix is positive definite.} 
\label{contours}
\end{figure}
\end{center}

$A$, $B$, $C$, $D$, $E$ involve Matsubara sums over bosonic frequencies $\omega_n=(2n+1)\pi T$ and 3-d integration,
which is done numerically. Analytic expressions can be obtained in weak coupling, using BCS hierarchy
of scales, $\delta\mu, \Delta << \mu$ (integration in thin shell around $\mu$). Concluding, in the weak coupling
gapless phases, $\delta\mu>\Delta$, are unstable (for both Goldstone and Higgs mode $B<0$, $E<0$, $C<0$). In the weak
coupling, the first order phase transition superfluid to normal state takes place, at $\delta\mu=\Delta/\sqrt{2}$,
before the gapless state developes, $\delta\mu=\Delta$; i.e. for the window $\Delta/\sqrt{2}<\delta\mu<\Delta$
Higgs and Goldstone modes are fluctuations around the meta-stable solution. To find stable gapless
phases one should explore the strong coupling regime.   

\begin{center}
\begin{figure}[h]
\includegraphics[width=2.5in,angle=0]{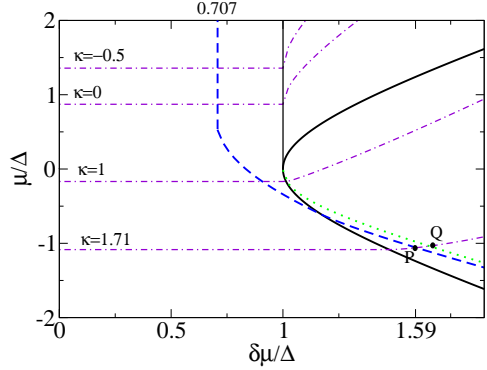}
\caption{The four dot-dashed lines show how $\mu$ varies as function of $\delta\mu$
for four different values of the dimensionless variable $\kappa=\pi/2\sqrt{2m\Delta} a_s$.
Values $\kappa<<-1$, $a_s<0$, correspond
to being deep in the BCS regime, while $\kappa>>1$, $a_s>0$, deep in the BEC regime. 
Region between curves 2 and 3, local and global minimum of free energy,
correspond to meta-stable gapless state.
Moving along $\kappa=1.71$ (BEC), point P correspond to the largest $\delta\mu$ where superfluidity
is globally stable, $\delta\mu/\Delta\approx 1.59$; after point Q, $\delta\mu/\Delta\approx 1.66$,
it is locally unstable.   
Curves 2 and 3  run closer as $\delta\mu$ increases, and converge asymptotically
for $\delta\mu>>\Delta$. Higgs mass is zero along curve 2. Hence Higgs mass is small along curve 3
and gets smaller as $\delta\mu$ increases.} 
\label{contours3}
\end{figure}
\end{center}

\section{Analysis of stability at $T=0$.}

Using the quasiparticle dispersion law, we identify regions in ($\delta\mu,\mu$) where the system has
gapless excitations, i.e. momenta where energy is zero. Gapless fermionic excitations live at one and two
spherical surfaces in momentum space. We map regions with gapless excitations and regions where coefficients $A$, $B$ etc.
are negative, indicating instabilities to the ($\delta\mu,\mu$) space.

Stability criteria are\\
1. Real speed of sound for Goldstone and Higgs modes ($B, E>0$),\\
2. SF is a local minimum ($C>0$),\\
3. SF is a global minimum ($\Omega_s-\Omega_n<0$),\\

which are equivalent to\\
1. Real speed of sound is equivalent to positive screening masses,\\
2. Local minimum is equivalent to positive number susceptibility.\\

At the Fig. \ref{contours}, the ($\delta\mu,\mu$) plane is divided into three regions
corresponding to different number of gapless surfaces in momentum space; also
unstable regions according to criteria 1 and 3 are shown.

Requirement 1 renders all states above the dotted and dashed lines
unstable, which leaves gapless states with one and two Fermi surfaces as possiblity. 
Requirement 3 renders all states above solid line unstable.

At the Fig. \ref{contours}, unstable regions according to criteria 2 and 3 are shown.

\begin{center}
\begin{figure}[h]
\includegraphics[width=2.5in,angle=0]{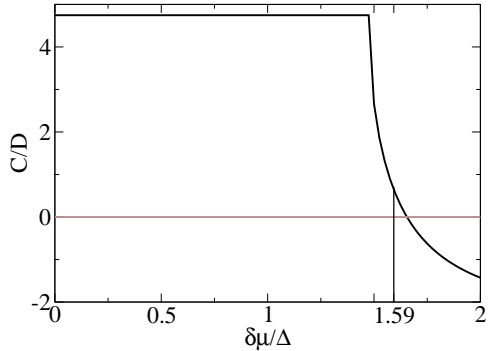}
\caption{Higgs mass, $m_H^2=C/D$, along $\kappa=1.71$
($g=1/k_Fa_s\approx 1.31$) curve as $\delta\mu$ increases.
For small $\delta\mu$, 
the Higgs mass is constant in the gapped superfluid phase. Indeed, in the gapped phase free energy
of superfluid is independent of $\delta\mu$, superfluidity is favored over normal and 
it is locally stable ($C>0$), hence $m_H$ is independent of $\delta\mu$. 
In gapless phase, Higgs mass decreases. Superfluidity wins over the normal, hence
the smallest $m_H$ in this regime ($C$ describes oscillations about the global minimum).
Between $P$ and $Q$, there is meta-stable region; below $Q$, it is locally unstable ($C<0$, $m_H$ is imaginary).
From $\delta\mu=0$ to $P$, $m_H$ drops by a factor of $7.5$.} 
\label{contours4}
\end{figure}
\end{center}

However, we still learn from criteria 1 about the tendency of the system 
to go to another state. Real speed of sound for the Goldstone mode, which is equivalent 
that Meissner mass is real (or current instability), is more stringent than the real speed of sound 
of Higgs mode. Both criteria point towards a non-homogeneous phase. We learn, that LOFF like
state is possible only in weak coupling at the BCS side. 

The global minimun of free energy is the strongest
requirement. There is a sliver of parameter space corresponding to meta-stable superfluid state  
(it is a local minimum, according to 2, but not a global one of free energy, according to 3).
Gapless superfluid state with one spherical Fermi surface is at BEC side ($\mu<0$).

\begin{center}
\begin{figure}[h]
\includegraphics[width=2.5in,angle=0]{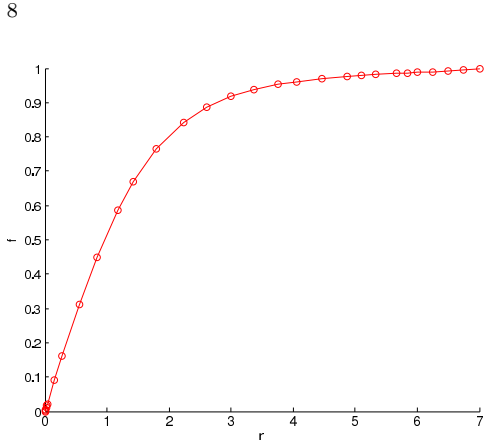}
\caption{Numerical solution of the ODE, Eq. \ref{ODE}, for the condensate in a vortex configuration.
Vortex center is at $r=0$. $\Delta(r)$ increases from $0$ to $\Delta$, as we go away from the inner core 
of the vortex, over a length scale $r_0=\sqrt{E/3C}$. Units are in $\Delta$ at both axis.} 
\label{contours5}
\end{figure}
\end{center}

\section{Parameters of the Higgs Lagrangian}

To be more concrete, we solve the gap equation for various scattering lengths and see where we land
in the parameter space. At the Fig. \ref{contours3}, the four dot-dashed lines show 
how $\mu$ varies as function of $\delta\mu$
for four different values of the dimensionless variable $\kappa=\pi/2\sqrt{2m\Delta} a_s$.
Region between curves 2 and 3, local and global minimum of free energy,
correspond to meta-stable gapless state.
Moving along $\kappa=1.71$ (BEC), point P correspond to the largest $\delta\mu$ where superfluidity
is globally stable, $\delta\mu/\Delta\approx 1.59$; after point Q, $\delta\mu/\Delta\approx 1.66$,
it is locally unstable.   
Higgs mass is zero along curve 2. Hence Higgs mass is small along curve 3
and gets smaller as $\delta\mu$ increases. Thus, there is a light Higgs in the gapless BEC superfluid state.

At the Fig. \ref{contours4}, we depict Higgs mass, $m_H^2=C/D$, along $\kappa=1.71$
curve as $\delta\mu$ increases.
For small $\delta\mu$, 
the Higgs mass is constant in the gapped superfluid phase.  
In gapless phase, Higgs mass decreases.
Between $P$ and $Q$, there is meta-stable region; below $Q$, it is locally unstable ($C<0$, $m_H$ is imaginary).
From $\delta\mu=0$ to $P$, $m_H$ drops by a factor of $7.5$.  

At point $P$:\\
a. At $P$, there is the smallest Higgs mass in the regime where superfluidity is favored over normal state.\\ 
b. Note about reliability of mean field. At point $P$, $g=1.31$ and as $\kappa$ increases, $g$ increases
even further. In BCS $g<0$ (large), BEC $g>0$ (large), since $g=1/k_Fa_s$ large $g$ corresponds to small $a_s$,
which means that the mean field is reliable.\\
c. Effective theory near $P$: gapless fermions living on one spherical surface in momentum space,
massless fluctuations in the phase of the condensate and massive but light fluctuations 
in magnitude of condensate. It is interesting to probe the spectrum.\\
d. Quantum corrections may alter small Higgs mass (beyond mean field).\\

It is interesting to explore part c., but even before a detailed study there is another striking
consequence of our results.

\begin{center}
\begin{figure}[h]
\includegraphics[width=2.5in,angle=0]{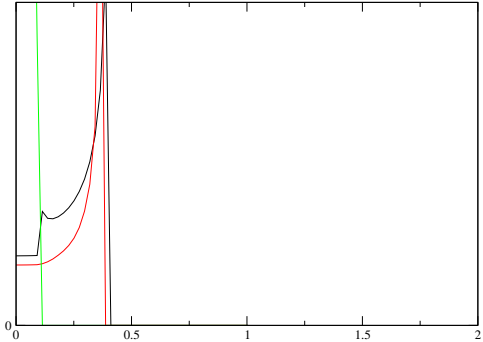}
\caption{Square of vortex radius $r_0^2$ as a function 
of a distance R from the center of a trap ($R\sim\delta\mu$). 
Two slightly different curves correspond to two ways of calculation. 
Bottom, smooth curve, first 3-d momentum integrals are taken 
numerically followed by taking a limit of small momentum. Top, curve with a casp,
the order of integration and small momentum limiting procedure is interchanged, which is
apparantly wrong.} 
\label{contours6}
\end{figure}
\end{center}

\section{Vortex in a trap}

The correlation length $r_0$, or the typical length scale at which
the magnitude of condensate $\Delta$ varies in field configurations that arise when the system is excited,
is inversely proportional to the mass of the Higgs mode of the system, $r_0~1/m_H$. 
For example, $r_0$ governs the size of the outer core of a vortex configuration in a superfluid phase.
This can be seen from the classical field equations for $\Delta(r)=f(r)e^{i\phi(\varphi)}$, $f(r)=\Delta+\rho(r)$,
for static configuration, where $\Delta$ is the ground state
value and $\rho(r)$ is fluctuation of the condensate, ($r,\phi$) are cylindrical polar coordinates,
and the vortex is at r=0,
\begin{eqnarray}
&& \Delta(r)-r_0^2\nabla^2\Delta(r) =const\nonumber\\
&& f(r\rightarrow 0 )=0,~~ f(r\rightarrow \infty)=\Delta
\label{ODE}
\end{eqnarray}
ODE is solved numerically, Fig. \ref{contours5}, with boundary conditions: zero condensate
at the center of the core, and it tends to a constant ground value at infinite $r$. 
For a vortex configuration, $\phi$ winds around by a multiple of $2\pi$ as we traverse 
a loop around the vortex. $\Delta(r)$ increases from $0$ to $\Delta$, as we go away from the inner core 
of the vortex, over a length scale $r_0=\sqrt{E/3C}$. The fact that $C$ (Higgs mass) is numerically small
close to the point $P$ in parameter space, will manifest itself in an increased size for the outer core
of the vortex.

To construct an actual vortex solution, it will be important to include the fourth order term 
in the effective action. However, the coefficient of this term is dimensionless and does not introduce
a new scale, hence our basic arguments remain valid.    
   
To see the effect quantitaively, we plot square of vortex radius $r_0^2$ as a function 
of a distance R from the center of a trap (harmonic trap in BEC; distance from the center of the trap
is proportional to mismatch in chemial potentials, $R\sim\delta\mu$). 
At the Fig. \ref{contours6}, in the gapped region 
the outer radius of the vortex is constant, and in the gapless region it increases monotonically
on increasing R (or $\delta\mu$). $r_0^2$ formally diverges at some point in the exterior of the trap,
which corresponds to the transition from superfluid to normal state. In reality this divergency 
is absent, since there are no vortices in the normal state, i.e. our formula does not apply then.
Two slightly different curves correspond to two ways of calculation. 
At the bottom smooth curve, 3-d momentum integrals are taken 
numerically first followed by taking a limit of small momentum. At the top curve with a casp,
the order of integration and small momentum limiting procedure is interchanged, which is
apparantly wrong. Here the long wavelength limit is not well defined, 
hence the order of operations matters. 

We estimate $r_0$ from the fact that it is the length scale at which the condensation energy is comparable
to the kinetic energy of a superfluid element close to the vortex, $E_k=E_{cond}$.
Taking into account the velocity
of superfluid matter near a vortex, which diverges towards the center of a vortex $v=\frac{1}{2mr}e_{\theta}$, we estimate
the kinetic energy, $E_k = n\frac{mv^2}{2}$. Condensation energy is proportional to a condensation energy density,
$E_{cond} = n\varepsilon_{cond}$. We get for the vortex radius $r_0\sim 1/\sqrt{\varepsilon_{cond}}$. Therefore,
in the gapped phase, $r_0$ is constant, since condensation energy density does not change;
in the gapless phase condensation energy density decreases with increasing $\delta\mu$, hence
$r_0$ increases steeply in the gapless phase; $r_0$ diverges at the transition 
from superfluid to normal state, because this formula does not work in the normal phase 
(there are no vortices). In conclusion, $r_0$ continuously increases in the gapless phase.
Vortex size grows steeply by entering into the gapless phase.

We suggest to tune parameters of the trap,  
the number of particles of the two species, $N_1$ and $N_2$,
and the scattering length, $a_s$, so that there is a sufficiently wide region in position space 
where the atomic system is in the gapless BEC phase. We predict, that effect of sharp increase
in vortex size can be seen. Probably, it requires  very flat traps to realize this phenomena in a wide
enough region to be observed.     

\section{Conclusion}

We obtained an effective theory for collective modes, Goldstone and Higgs, for the system
of interacting two fermionic species. There is an interesting low energy content of effective theory,
which includes gapless fermions, massless Goldstone mode, and very light Higgs mode. From our phase diagram,
gapless state is stable at the BEC side of the resonance, with
gapless fermions residing on one spherical Fermi surface in momentum space. Instability towards
non-homogeneous LOFF state occurs only at weak coupling, i.e. LOFF is favored away from the resonance. 
 
We would like to study the low energy effective theory further. Apart from that, we suggest 
a possible experimental evidance/signature of the gapless phases. We predict a sharp increase
in outer core of a vortex, i.e. vortex size, upon entering into the gapless phase. To observe
this dramatic effect will require tuning of the parameters of the trap to a gapless BEC phase.

It will be interesting to study the core structure of the vortex,
including inner and outer core, as well as interaction between two vortices. This requires
solving Bogoluibov-De Ginnas equation and/or constructing the Landau-Ginsburg functional
to the fourth order.
However, we believe that our prediction with respect to vortex size will not change qualitatively.
According to our estimate, the mean field treatment is reliable in the region of interest.
Probably, the main improvement will come from incorporating momentum/energy dependent 
Fermi interaction.

\end{document}